# Consolidating a Model for Describing Situated Software Practices

Diana Kirk
*Technology Academy, EDENZ Colleges, 85 Airedale Street, Auckland 1010, New Zealand*

Stephen G. MacDonell
*SERL, School of Engineering, Computer & Mathematical Sciences, Auckland University of Technology, Private Bag 92006, Auckland 1142, New Zealand*

Ewan Tempero
*Department of Computer Science, University of Auckland, Private Bag 92019, Auckland 1142, New Zealand*

**Abstract**

*Many prescriptive approaches to developing software intensive systems have been advocated but each is based on assumptions about context. It has been found that practitioners do not follow prescribed methodologies, but rather select and adapt specific practices according to local needs. As researchers, we would like to be in a position to support such tailoring. However, at the present time we simply do not have sufficient evidence relating practice and context for this to be possible. We have long understood that a deeper understanding of situated software practices is crucial for progress in this area, and have been exploring this problem from a number of perspectives. In this position paper, we draw together the various aspects of our work into a holistic model and discuss the ways in which the model might be applied to support the long term goal of evidence-based decision support for practitioners. The contribution specific to this paper is a discussion on model evaluation, including a proof-of-concept demonstration of model utility. We map Kernel elements from the Essence system to our model and discuss gaps and limitations exposed in the Kernel. Finally, we overview our plans for further refining and evaluating the model.*

## 1. INTRODUCTION

The history of software engineering process has been one of *advocacy*. Many authors, from both academia and industry, have architected software development methodologies and processes in the belief that strict adherence by practitioners will inevitably result in project success. Examples include traditional models such as waterfall (Royce, 1970) and spiral (Boehm, 1988) and agile methods such as eXtreme Programming (XP) (Beck, 2000) and Feature Driven Development (FDD) (Luca, nd). The wisdom manifested was that "development models are best regarded as a coherent set of practices, some of which are required to balance the performance trade-offs arising from the use (or absence) of others" (Cusumano et al., 2003).

However, each of the proposed methodologies is based on a number of assumptions about context. For example, testing in XP requires an active customer to test a delivered increment and provide feedback i.e. it is assumed that a capable and empowered customer is available and that the product has a human interface. Turk et al. point out that this is inappropriate if the product-under-development is, for example, an embedded system (Turk et al., 2005). Agile approaches in general tend to advocate continuous refactoring, but again this may be inappropriate when the system to be developed is large and complex. The notion of 'tailoring according to contexts' has become the accepted wisdom (MacCormack et al., 2012; Müller et al., 2009; Petersen and Wohlin, 2009; de Azevedo Santos et al., 2011; Turner et al., 2010). However, at the present time, we as researchers are not in a position to support such tailoring because we simply do not have sufficient evidence relating practice and context.

We have long been interested in deepening our understanding of the relationships between practice and context, and have been exploring this problem from a number of perspectives. Three key insights have been gained from these explorations. First, we cannot know exactly what a practice entails without an awareness of the context in which it was enacted i.e. we must now view practice and context as inherently intertwined. This notion has precedence in the organisational and management sciences (Orlikowski, 2002) and our recent research has exposed this dependence for software practices (Kirk and MacDonell, 2016). Second, if we wish to discuss and compare disparate practices, we must apply a research frame- work that focusses on *what* a practice is meant to achieve, rather than *how* it achieves it (Kirk and Tempero, 2012a). For example, an organisation producing medical equipment for a specific client may implement a strict requirements elicitation process involving formal sign-off by the client, whereas a startup organisation may adopt a 'try it and see' approach with feedback from many clients. Both approaches aim to establish requirements, but do so in very different ways. We also note that each alters the environment, but in a different way. In the former case, the product becomes more formally defined whereas in the latter case, understanding resides in the heads of the development team. Third, when exploring *con- text* in the literature, we found that many different *kinds* of contextual factor were mentioned. In addition to factors that described context at an operational level, we also found factors that described project objectives, factors that described high level organisational objectives and factors that were simply un- clear (Kirk and MacDonell, 2018; Kirk and Buchan, 2018). Factors tended to be organised in a way that does not comply with Nickerson et al.'s viewpoint that the criteria for classification must be identified and must not be purely descriptive in nature (Nickerson et al., 2013). For example, the framework proposed by Clarke and O'Connor contains a factor 'Cohesion' which includes 'team members who have not worked for you', 'ability to work with uncertain objectives' and 'team geographically distant'. There is no clear criteria upon which to base classification decisions as these sub-factors have different *meanings*.

Our previous explorations have led us to adopt the position that, before we can confidently support practitioners in their tailoring efforts, we must address two issues. First, we must build an evidence base that relates a situated practice (a practice performed within a specific context) and its outcomes with respect to desired objectives. As pointed out by Lengnick-Hall and Griffith, if the intention is to achieve "a specific, designated outcome", as is the case for most software practices, the knowledge (practice) must be applied as-is. Applying the knowledge in an intuitive or experimental way introduces a lack of fit between type of knowledge and how it is applied, and this inevitably leads to reduced effectiveness, at best (Lengnick-Hall and Griffith, 2011). From this perspective, ad-hoc tailoring that is not grounded in evidence might be viewed as a 'hidden' issue. Second, we require some notion of 'equivalent practices'. For example, we might view the two requirements elicitation practices mentioned above as 'equivalent', as each results in an understanding of scope i.e. they are similar as regards 'what' is achieved.

In this position paper, we draw together the models we have created to address the issues highlighted above into a holistic model. These models represent the exploratory stages of our research. We are now entering a second, more formal, refinement stage (Routio, 2007). This approach is consistent with the notion of developing theory (models) inductively with "inductive theory building ... producing new theory from data" followed later by "deductive theory testing completing the cycle by using data to test theory" (Eisenhardt and Graebner, 2007). We discuss the ways in which the model might be applied to sup- port the goal of evidence-based decision support for practitioners and overview our plans for further refining and evaluating the model. Contributions specific to this paper include a discussion on our approach to model evaluation and a critique of the established *Essence Kernel* based on a mapping between Kernel and model.

In section 2, we overview the model and in section 3 we discuss how it can be used to support evidence accumulation and (eventually) decision support. In section 4, we present our approach to evaluating and further refining the model and discuss the results from mapping the Essence Kernel onto our model. In section 5, we briefly overview related work and in section 6, we summarise the paper.

## 2. MODEL OVERVIEW

In this section, we present the model that is the result of previous works in which we explored various aspects of situated software practices (Kirk et al., 2009; Kirk and Tempero, 2012a; Kirk and Mac- Donell, 2014; Kirk and MacDonell, 2018; Kirk and MacDonell, 2016; Kirk and Buchan, 2018). We overview the model in Figure 1 and Table 1. At top level is a *Software Initiative*. A *Software initiative* is any endeavour that involves defining, creating, delivering, maintaining or supporting software intensive products or services. It thus encompasses the more re- cent client-focussed delivery paradigm and subsumes the traditional 'project'.

We view a software initiative as a set of *practices* implemented with the aim of meeting specified *objectives* and enacted within an *operational context*. This is in keeping with an earlier understanding that "Sound tailoring requires the ability to characterise ... goals ..., the environment ..., and the effect of methods and tools on achieving these goals in a particular environment" (Basili and Rombach, 1987). The initiative thus comprises some *objectives*, for example, 'time-to-market', a set of *practices*, for example, 'design review', and an *operational context*. The operational context is many-faceted and includes factors that *directly* affect practice efficacy, for example, cultural mis-matches between development teams.

The initiative exists within a higher level *strategic context*. Strategic context factors do not directly affect practice efficacy, but rather spawn decisions relating to *objectives* and *operational context* i.e. the effect is indirect. Examples are the organisation's need to gain consumer trust, or to expand into a global market- place. The first may result in a project where product quality is stated as the key objective. The second may result in the establishment of off-shore teams, thus affecting operational context.

*Operational context* abstracts the many factors that are of direct interest for practice efficacy. *Working context* comprises the base dimensions of our con- text mode i.e. the elements that must be taken into account for practice selection and tailoring. The four dimensions are:

Table 1. Local operational context factors.

| | | |
|---|---|---|
| People | Entity | Capability |
| | | Motivation |
| | | Empowerment |
| | | Team cohesion |
| | Interface | Team cohesion |
| Place | Entity | Physical distance |
| | | Temporal distance |
| | | Availability |
| | Interface | Physical distance |
| | | Temporal distance |
| | | Availability |
| Product | Product type | e.g. embedded |
| | Lifecycle stage | e.g. new, mature |
| | Standards | e.g. safety |
| | Requirements | e.g. clear, complete |
| | Implementations | e.g. consistency, modularity |
| Process | Client | |
| | Parent org | |
| | Legal | |
| | Financial | |

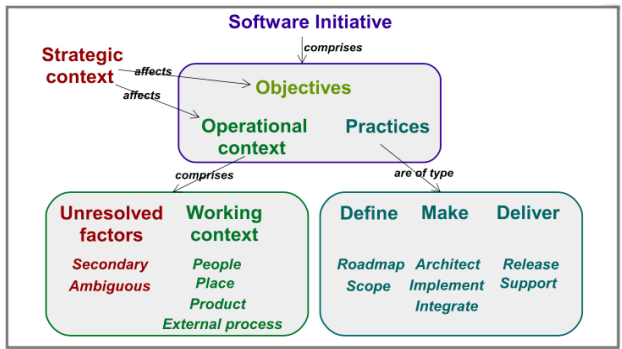

Figure 1: Model for situated software practices.

**People.** Cultural characteristics affecting peoples' ability to perform

**Place.** Peoples' availability affecting logistics and communications

**Product.** Characteristics of the product that is being developed

**Process.** Processes external to the initiative (as compared with practices within the initiative).

The sub-categories for *working context* are shown in Table 1.

However, our earlier investigations revealed that some of the terms commonly used in the literature are simply not sufficiently clearly defined and we con- sider these to be *Unresolved factors*. *Secondary* factors are those made up of multiple ideas. An example is 'outsourcing', which has many possible scenarios relating to *what (product)* is outsourced by *whom (people)*, *where (place)*. *Ambiguous* factors require deeper consideration. An example is 'uncertain requirements', which may exist because the client is un- clear about what is wanted (*product*), client processes result in delays before decisions can be made (*pro- cess*), or the client isn't available (*people*).

Our abstraction for *practices* resulted from an earlier study in which we wished to understand why some SMEs are successful, despite failing to follow the standard process reference models (Kirk and Tempero, 2012a). This led to our subsequent viewpoint that we will be in a position to compare practices only if we categorise based on *function*. Our functional categories are shown in Table 2.

Because we have structured based on *function*, the categorisation will support any practice deemed to be relevant for meeting objectives. For example, informal meetings in the lunch room that help developers understand scope clearly fit into the 'scope' category. Of note is that project management activities such as tracking are not included. Such activities do not directly address objectives, but rather impact *context* for example, team cohesion, and integrate into our model via the contexts affected. A precondition that an objective be met is that each category must contain one or more effective practices. To illustrate, for a 'Quality' objective, including quality considerations during product architecture, implementation and integration will fail to achieve the objective if quality expectations are not included during scoping. The identification of gaps in the overall process is straightforward.

Table 2: Categories for practices.

| | |
|---|---|
| **Define** | Roadmap Scope |
| **Make** | Architect |
| | Implement |
| | Integrate |
| **Deliver** | Release Support |

A characteristic of the above categorisation is that of *flexibility*. There is no assumption about ordering of practices, and no expectation that, for example, practices relating to *Implement* and *Integrate* need be separate. Such ordering would exist at a higher level and might be used to describe strategies of iteration and incremental delivery. We also submit that the categorisation is 'paradigm-agnostic' — whether an initiative is run in a traditional or agile way, the basic functions of defining, making and delivering the product must be carried out. Of note is the fact that the sub-categories 'Roadmap' and 'Support' lie out- side of a traditional development project and subcategories 'Define' and 'Deliver' relate to practices that span development organisation and client.

## 3. PROGRAMME OF RESEARCH

In this section, we overview our planned programme of research. Our overall goal is to support practitioners with selecting practices that are suitable for their local context. We note that studies carried out are expected to result in refinements to our research model as our understanding deepens (Routio, 2007). We also note that studies will act as an evaluation of the model (see section 4).

### 3.1 Practice Operating Limits

Our immediate plans relate to growing an evidence base for situated software practices. We will include studies to investigate within industry the indicated and contra-indicated contexts for specific practices. We will ask practitioners to select recently- or currently- enacted practices and elicit information about contextual factors that were perceived as supporting, and those perceived as detracting from, meeting objectives. The studies will take two forms. For some, we will use our model as a guide and in others we will elicit information from the practitioner perspective only. The goals of these studies are a) to refine and evaluate the model, and b) to accumulate evidence relating to 'happy' and 'unhappy' contexts for specific practices. This evidence will later be used as inputs to decision support mechanisms. We note that the usefulness of this evidence will be reliant on a context model that is complete i.e. evidence from early studies may be ineffective if the model changes too much during refinement. However, as mentioned above, we must investigate practice and context together as these are inherently intertwined - we cannot implement a set of studies on context alone.

### 3.2 Families of Practices

For this avenue of research, we want to build up 'families' of 'equivalent' practices. The aim is to suggest to the practitioner a set of practices that per- form the same function. For example, in section 1, we discussed two possible ways of eliciting requirements (*product scoping*), one involving formal documentation and the other involving a 'try it and see' approach. A major challenge in this research will be the establishing of exactly what we mean by 'equivalent'. Certainly, the starting point is with the function to which the practice belongs ('Scope'). However, implementation of a practice has the side-effect of changing context in some way, and it is unclear at this point how this affects things, if at all. Once the research has matured, we would expect to use the notion of equivalence to suggest possible alternatives to practitioners during decision support.

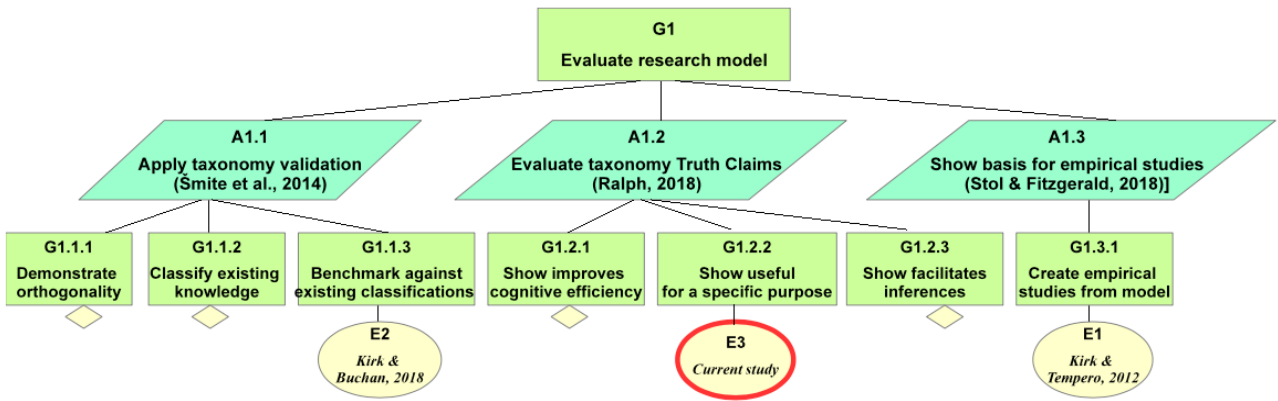

Figure 2: Evidence tree for model.

### 3.3 Decision Support

Once we are sufficiently confident in our model, we plan to use this along with evidence accumulated to create a decision support mechanism for supporting practitioners with practice tailoring. As there are many kinds of software initiative and many possible situational factors, we envisage offering recommendations based on available evidence, leaving final decisions to be made by practitioners with local knowledge. Clearly, an extensive body of evidence must be accumulated before this can be implemented.

## 4. EVALUATION

For software engineering, it is suggested that an appropriate approach to validation is to gather different kinds of evidence from various sources. Dawson et al. make a case for accumulating many little evidences of various kinds (Dawson et al., 2004). Others suggest the researcher should accumulate an 'evidence portfolio' containing, for example, case studies, anecdotes, surveys, expert opinion and controlled experiments (Kitchenham et al., 2005; Weaver et al., 2005). We adopt this approach for our research model. We implement a technique called *argumentation* and capture evidence goals, arguments and evidence items using *Goal Structuring Notation (GSN)* (GSN Working Group, 2018). The notation helps the researcher see what evidence is missing and allows stakeholders to see 'at a glance' what is the evidence coverage.

The evidence tree for our model is shown in Figure 2. Our top level goal is 'evaluate research model'. We address this goal by sourcing viewpoints on model/theory/taxonomy validation from the literature. Smite et al. suggest that a taxonomy must exhibit dimensional orthogonality, should be used to classify existing knowledge and should be benchmarked against existing classifications (Smite et al., 2014). These are shown as argument strategy A1.1 and sub-goals G1.1.1, G1.1.2 and G1.1.3. Ralph de- fines three 'truth claims' inherently made by a taxonomy and suggests these form a suitable set for evaluation (Ralph, 2018). These are shown as A1.2 and G1.2.1, G1.2.2 and G1.2.3. Stol and Fitzgerald re- quire that a model can be used as a basis for empirical studies (Stol and Fitzgerald, 2018). These are shown as A1.3 and G1.3.1.

As is clear from the figure, only two pieces of evidence have thus far been addressed in earlier work (Kirk and Tempero, 2012b; Kirk and Buchan, 2018). We are in the process of implementing two further studies, the first to elicit practitioners' perspectives on context, and the second to explore the operating limits for two agile practices. The first will address goal 1.1.2 (classify existing knowledge) and the second goal 1.3.1 (create empirical studies from model). In the next section, we describe a mapping that addresses goal 1.2.2. (show useful for specific purpose).

The approach is useful for exposing *breadth* of evidence but those interested must read the research out- puts to establish the *quality* of the provided evidence.

### 4.1 Essence Analysis

In this section, we carry out a 'proof-of-concept' map- ping between the *Essence Kernel* from the SEMAT initiative (Object Management Group, 2018) and our research model. The SEMAT initiative aims to pro- vide a theoretical basis to support comparison, evaluation and tailoring of software practices. Central to the scheme is a *Kernel*. Practices are described in *Kernel elements*, which are "integral to all software engineering methods". The kernel comprises *Alphas* (attributes relating to assessing project health), *Activity spaces* (the essential things to do) and *Competencies* (key capabilities required). Alphas capture the main software engineering concepts along with pre-defined states and checklists, thus providing critical indicators for the measurement of project health. Activity spaces provides a partitioning of the 'things to do' i.e. addresses the activities that must be carried out. Competencies provide a view of the key capabilities required to carry out activities.

As the aim of the *Kernel* is similar to our aims i.e. to support practitioners in adapting methods in an agile way to suit their needs, we felt it necessary to attempt a mapping between the two models as a means of testing that our abstraction is sufficiently comprehensive to allow us to critique an established approach. In this section, we map the elements of the *Kernel* to our model and discuss findings. The results of the mapping are shown in Table 3.

According to our understanding, one of the Alphas (*Opportunity*) maps to both *Strategic context* and an aspect of *Working context*, two map to practice- related ideas and the remaining four to *Working con- text*. As these seem to include different kinds of entity, we checked the possible states for each Alpha to gain a deeper understanding. The first point of interest is found in the mappings of *Requirements* and *Soft- ware system* to *Product*. States for *Requirements* include 'Bounded', 'Coherent' and 'Fulfilled' and these are certainly consistent with our notions of 'Clear' and 'Complete' (see Table 1). However, *Software system* in the *Kernel* is exemplified by, for example, 'Usable', 'Ready' i.e. describes the application with respect to readiness (a project management perspective), whereas our abstraction describes the product characteristics we believe to be relevant for practice tailoring, for example, the degree of consistency be- tween representations, design modularity, etc.

For *Activity spaces* as might be expected, most of the *Kernel* elements map almost directly to *Practices* in our model. However, our model contains no elements for the project management oriented activities such as 'Support the team'.

For *Competencies*, again as expected, most items map to *People:Entity:Capability*. However, the *Kernel* defines the set of competencies required in a soft- ware project whereas, in our model, the 'capability' attribute represents the level of capability of a team with respect to a specific practice.

*4.1.1 Discussion*
The two approaches are very different as regards abstraction choice. The *Kernel* perspective appears to be more project management oriented, with Alphas including notions of all of *Who* (people), *What* (product) and *How* (process). Our perspective tends to be more focused on 'what does a team really need to know to adapt practices?' i.e. the perspective is that of the people doing the work. The *Kernel* elements with no mapping in Table 3 represent project management activities. As discussed in section 2, these activities affect objectives indirectly, via working context. To illustrate, rather than knowing that someone is 'supporting the team' or 'coordinating activities', we need only know to what extent the team feels empowered and cohesive as this affects tailoring choices. A second observation relates to the pre-defining of Alpha states. First, the inference is that these states are relevant for *all* software initiatives. However, it is not clear how the *Requirements* state of 'Fulfilled', i.e. that the requirements addressed "... fully satisfy the need for a new system", apply in the very common situations of the team and/or the customer being un- sure about what is wanted i.e. when an evolutionary approach is appropriate, perhaps in a startup situation. Second, we do not need to know that a team is in state 'formed', we actually need to know if members share a room or are spread across different countries, as this will directly impact on which practices might be effective. We could not find such context-related ideas represented in the *Kernel*.

The fact that most of the *Kernel* competencies map to *People:Entity:Capability* illustrates the differences in the viewpoint of 'prescribe what competencies are required' and 'whatever practice is being considered, we need to know the competency level of the team with respect to the practice'. Our critique of the *Kernel* approach is the implicit assumption that the complete set of required competencies is known i.e. the approach does not support extension. For example, a startup group must be innovative, a competency not currently included in the *Kernel*.

Another consideration relates to the notion of *flexibility*, In agile approaches, developers often combine, for example, design and implementation. Activity spaces in the *Kernel* appear to be fixed and disjoint and it is not clear that such flexibility is supported.

In section 1, we presented our viewpoint from earlier studies that, before we are in a position to offer decision support for tailoring, we must first a) gather evidence relating context, software practice and out- comes, and b) understand practice 'equivalence' in or- der that we know when it is acceptable to replace one practice with another and can identify gaps in the pro- cess. We cannot see how the *Kernel* is able to fully address either of these. A detailed notion of relevant context is a pre-requisite for the first. The second implies that the set of practices for an objective must include an element from each category. There is no point in implementing strong quality related practices during architecture, implementation and release if the 'scoping' practices are weak from a quality perspective. The statement in section 8.1.1 that the *Kernel* allows you "... to apply as few or as many practices as you like" is not consistent with the notion of identifying practice gaps.

Finally, we notice that none of the *Kernel* elements maps to our *Working context* dimensions *Place*, *Product* or *Process*. We have identified these as being crucial ideas for selecting and tailoring practices. Some examples of different practices being indicated include co-located versus dispersed team members, (un)availability of architectural documentation and the culture of the parent organisation. Our main critiques relates to the lack of any deep consideration of working context, including type of product and locational information. Without this, any tailoring can not be grounded in evidence and remains potentially problematic, as discussed in section 1.

# 5. RELATED WORK

Routio describes three kinds of research as a) there is no model to use as a starting point (exploratory re- search), b) an existing model is being expanded or refined, and c) hypotheses based on an established model are being tested. Exploratory research is ap- propriate for a "phenomenological pursuit into deep understanding". The researcher begins with a "pre- liminary notion" of the object of study. During the study the "provisional concepts ... gradually gain precision" until a suitable conceptualisation is achieved (Routio, 2007). We have adopted this approach for our research.

Table 3: Mapping between Essence Kernel and model.

| Alphas | Opportunity | Reason for creation of ... system | Strategic context |
|---|---|---|---|
| | | Shared understanding of stakeholders' needs | People::Capability |
| | Stakeholders | People, groups or organizations affected ... | People |
| | Requirements | What the software system must do | Product:Req |
| | Software system | System of software, hardware and data | Product:Impl |
| | Work | Activity to achieve a result | Practices |
| | Team | People engaged in implementation activities | People |
| | Way-Of-Working | Tailored set of practices to support work | Practices |
| Activity Space | Explore possibilities | Opportunities to be addressed | Practices::Roadmap |
| | | Identify stakeholders | Practices::Roadmap |
| | | Understand stakeholder needs | Practices::Scope |
| | Ensure stakeholder satisfactn | Gain acceptance from stakeholders | Practices::Release |
| | Use the system | Observe benefits in live environment | Practices::Support |
| | Understand the requirements | Gain shared understanding of products | Practices::Scope |
| | Shape the system | Shape for reuse and expected demands | Practices::Architect |
| | Implement the system | Implement and test system elements | Practices::Implement |
| | | Build by integrate and test system elements | Practices::Integrate |
| | Deploy the system | Make system available externally for use | Practices::Release |
| | Operate the system | Support system in a live environment | Practices::Support |
| | Prepare to do the work | Set up team and working environment | No mapping |
| | Coordinate activity | Co-ordinate and direct team's work | No mapping |
| | Support the team | Support team to help themselves | No mapping |
| | Track progress | Measure and assess progress | No mapping |
| | Stop the work | Shut down SE endeavour | No mapping |
| Competencies | Stakeholder representatn | Team understands customer needs | People::Team cohesn |
| | Analysis | Ability to transform needs to requirements | People::Capability |
| | Development | Ability to design and program SW systems | People::Capability |
| | Testing | Ability to verify system meets requirements | People::Capability |
| | Leadership | Ability to motivate and inspire a team | People::Capability |
| | Management | Ability to coordinate, plan and track work | People::Capability |

Ralph suggests that a framework or model may be viewed as a taxonomy and represents a theory stated as "a system of ideas for making sense of what exists ... in a domain." He suggests that a taxonomy by its nature posits three 'truth claims', a) the taxonomy improves cognitive efficiency, b) it facilitates inferences and c) it is useful for a specific purpose. Evaluation involves addressing each of these claims (Ralph, 2018). We have included these truth claims as items in our evidence tree.

Stol and Fitzgerald present a set of metaphors for the various kinds of research within software engineering (Stol and Fitzgerald, 2018). The metaphor applied for building theories or models with some degree of generalisability is that of a *jigsaw puzzle*. Each piece of the puzzle is considered and theorised with respect to the whole. The authors suggest that such studies are important as they provide "a new conceptual lens to design future empirical studies". Resulting models must, of course, be validated during sub- sequent empirical studies. The refinement stage for our context model fits into this category, as we are effectively uncovering new pieces of the puzzle and integrating into the whole.

## 6. CONCLUSIONS AND FUTURE WORK

In this position paper, we have drawn together the results of our previous studies to create a holistic model which we plan to use as a basis for accumulating evidence relating to situated software practices. Such evidence is a pre-condition for our long term goal of supporting evidence-based decision making for practitioners who wish to tailor software practices to suit local context.

Contributions specific to this paper include a discussion on our approach to model evaluation and a critique of the established *Essence Kernel* based on a mapping between Kernel and model. The mapping exposed several issues relating to generality and flexibility and rooted in the pre-defining of *Alpha states*, *Competencies* and *Activities*. We conclude that, even at this early stage, our conceptual model has proven useful.

The model is now entering a 'use-and-refine' cycle. Future work planned includes a) consolidating the context-related elements of the model, and b) ac- cumulating evidence on operating limits for specific situated software practices.